\documentclass[preprint,prd,aps,onecolumn,english]{revtex4}
\usepackage{amsmath,amssymb} 
\usepackage[dvips]{graphicx} 
\usepackage{graphics}
\usepackage{graphicx}
\usepackage{epsfig}
\usepackage[english]{babel}
\usepackage[cp1251]{inputenc}

\newcommand\nn{\nonumber}
\newcommand\ba{\begin{eqnarray}}
\newcommand\ea{\end{eqnarray}}

\begin{document}

\title{The decay $\tau \to (\pi, \pi') \nu_{\tau}$ in the Nambu - Jona-Lasinio model}
\author{A.~I.~Ahmadov$^{a,b}$ \footnote{E-mail: ahmadov@theor.jinr.ru},
M.~K.~Volkov$^a$ \footnote{E-mail: volkov@theor.jinr.ru}}
\affiliation{$^{a}$ JINR-BLTP, 141980 Dubna, Moscow region,
Russian Federation}
\affiliation{$^{b}$ Institute of Physics, Azerbaijan
National Academy of Sciences, H.Javid ave. 131, AZ-1143 Baku, Azerbaijan}

\date{\today}

\begin{abstract}
In the Nambu - Jona-Lasinio model the decay widths $\tau \to (\pi, \pi'(1300)) \nu_{\tau}$
were calculated.
For the calculation of the decay width $\tau \to \pi \nu_{\tau}$ the standard NJL model was used with taking
into account the intermediate axial-vector meson and the extended Nambu - Jona-Lasinio model with the intermediate
$a_1(1260)$ and $a_1'(1647)$ mesons. These results are in good agreement with each other.
For the calculation of the decay width $\tau \to \pi' \nu_{\tau}$ extended Nambu - Jona-Lasinio model was used.
The results are in satisfactory agreement with the experimental data.

Numerical estimates for the constant $F_{\pi}$ and $F_{\pi'}$ are also  in satisfactory agreement with both the experimental values for
$F_{\pi}$ and the results obtained by calculations on the lattice.

\vspace*{0.5cm}
\noindent
PACS: 11.30.Rd; 13.20.-V; 13.25.-K
\end{abstract}

\maketitle

\section{Introduction}
\label{Introduction}
In the recently fulfilled works \cite{volkov1,volkov2,volkov3,volkov4} a number of decays of the $\tau$ leptons were described.
The intermediate processes with the participation of vector mesons in both ground and radial excited states
were taken into account. For this, the extended NJL model was used \cite{volkov5,volkov6,volkov7,volkov8}.
This model allows us to describe all these processes without introducing any additional arbitrary parameters
unlike the methods of other authors \cite{paver1,paver2,Guo,Roig}.
The decay widths $\tau \to \pi^- \pi^0 \nu_{\tau}; \,\,\,\,\tau \to \pi \eta (\eta')\nu_{\tau};\,\,\,\,
\tau \to 2\pi \eta (\eta')\nu_{\tau};\,\,\,\,\tau \to \pi \omega \nu_{\tau}$ were calculated .

In this work, we will describe one of the main $\tau$ lepton decays $\tau \to \pi \nu_{\tau}$ and give the estimate
for the decay width $\tau \to \pi'(1300) \nu_{\tau}$.

The calculation of the decay width $\tau \to \pi \nu_{\tau}$ was conducted in both the standard and in the framework extended NJL models.
These results are in good agreement with each other.
Calculations of the decay width $\tau \to \pi'(1300) \nu_{\tau}$ were made in the extended NJL model.
The estimates obtained for the decay widths $\tau \to \pi(\pi'(1300)) \nu_{\tau}$ are in satisfactory agreement with the experimental data
\cite{pdg,exp}.
The value obtained for $F_{\pi'} =$ 4.68\,\,NeV is in a qualitative agreement with the estimate obtained by calculation on the lattice \cite{Kim}.

\section{Lagrangians of quark-meson interactions}

In the extended NJL model, interactions of the $\pi, \,\,\pi'(1300)$ and $a_1(1260),\,\, a_1' (1647)$ mesons with quarks
are of the form:
\begin{eqnarray}
\label{eq:l}
&&\Delta\mathcal{L}^{int}_E(q,\bar{q},a_1,a_1',\pi,\pi') = \\ \nonumber &&\,\,\,=\bar{q}(k') \left[ \left(A_1 a_1^{(-)\mu}\right.\right.
- \left.\left.   A_2{a_1'}^{(-)\mu} \right)\tau^{-} \gamma^\mu \gamma^5\right. +
\left.  \left( P_1 \pi^- - P_2\pi'^-\right) \tau^{-} \gamma^5 \right]q(k), \\ \nn
&&\tau^- = \sqrt{2}\left(\begin{array}{cc} 0&0 \\1&0 \end{array} \right),
\end{eqnarray}
where $a'$ and $\pi'$ are the axial-vector and pseudoscalar meson fields in the excited state,
\begin{eqnarray}
\label{eq:2}
A_1 &=& \frac{g_{\rho_1}}{2} \frac{\sin(\beta+\beta_0)}{\sin(2\beta_0)}+ \frac{g_{\rho_2}}{2} f(k_\bot^2)\frac{\sin(\beta-\beta_0)}{\sin(2\beta_0)},\\ \nonumber
A_2 &=& \frac{g_{\rho_1}}{2} \frac{\cos(\beta+\beta_0)}{\sin(2\beta_0)}+\frac{g_{\rho_2}}{2} f(k_\bot^2)\frac{\cos(\beta-\beta_0)}{\sin(2\beta_0)},\\ \nonumber
P_1 &=& g_{\pi_1} \frac{\sin(\alpha+\alpha_0)}{\sin(2\alpha_0)}+g_{\pi_2} f(k_\bot^2)\frac{\sin(\alpha-\alpha_0)}{\sin(2\alpha_0)},\\ \nonumber
P_2 &=& g_{\pi_1} \frac{\cos(\alpha+\alpha_0)}{\sin(2\alpha_0)}+g_{\pi_2} f(k_\bot^2)\frac{\cos(\alpha-\alpha_0)}{\sin(2\alpha_0)}.
\end{eqnarray}
The values of the angles $\beta=79.85^\circ$ and $\beta_0=61.44^\circ$ are taken from \cite{volkov6},
and the angles $\alpha=59.38^\circ$ and $\alpha_0=59.06^\circ$ correspond to the mixing angle taken from the vector meson sector \cite{volkov10}.

The form factors $f(k_\bot^2)$ used in the extended NJL model have the form $f(k_\bot^2)=(1+d k_\bot^2)$, where
$k_\bot =k-\frac{(kp)p}{p^2},$ is the transverse quark momentum, for light quarks $d=-1.784~\mathrm{GeV}^{-2}$ \cite{volkov6}.
Here $k$ and $p$ are the quark and meson momenta, respectively.
In the meson rest frame, this formula takes the simple form $k_\bot =(0,\vec{k}).$  \\
$\pi$ and $\pi'$ are the pion fields in the ground and first radial excited states. \\
$a_1$ and $a_1'$ are the axial-vector meson fields in the ground and first radial excited states.  \\
$q$ is the field of the $u$ and $d$ quarks and $\bar {q} = (\bar {u}, \bar {d})$. \\
The quark-meson coupling constants are
\begin{eqnarray}
g_{\rho_2} = \left(\frac23 I_2^{f^2}(m_u)\right)^{-1/2},\qquad g_{\pi_2}=\left(4 I_2^{f^2}(m_u)\right)^{-1/2}, \nn \\
g_{\rho_1} = \left(\frac23 I_2^{(0)}(m_u, m_d)\right)^{-1/2}=6.14, \qquad
 g_{\pi_1} = g_{\pi} =\frac{m_u}{F_{\pi}},
\end{eqnarray}
where the integrals $I_m^{f^n}$ read
\begin{equation}
I^{f^{n}}_m(m_q) = -i \frac{N_c}{(2\pi)^4} \int \mbox{d}^4 k\frac{(f_q({k_\bot}^2))^n}{(m_q^2-k^2)^m}\Theta(\Lambda^2_3 -  k_\bot^2),
\end{equation}
and the cut-off parameter $\Lambda_3 = 1.03~$ GeV~\cite{volkov2,volkov6}.

Add here the Lagrangian describing a weak interaction of the lepton and quark current
\ba
\mathcal {L}^{weak} = \bar {\tau} \gamma_{\mu}\gamma_5 \nu \frac{G_F}{\sqrt{2}}|V_{ud}| \bar {d} \gamma_5 \gamma_{\mu} u.
\ea

These Lagrangians allow us to calculate the decay widths $\tau \to (\pi \pi' (1300)) \nu_{\tau}$ without introducing
any extra parameters.

\section{The decay $\tau \to \pi \nu_{\tau}$}

\begin{figure}[!htb]
       \centering
       \includegraphics[width=0.5\linewidth]{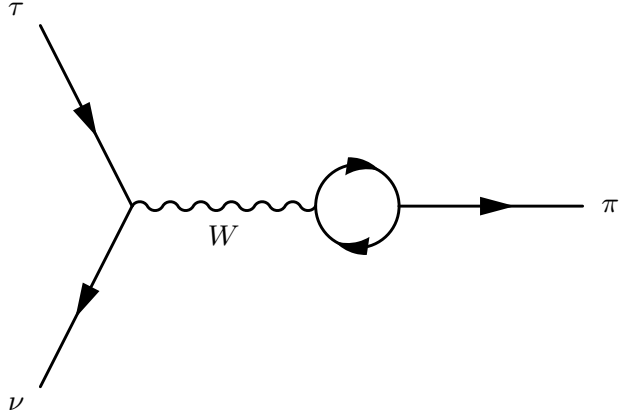}
       \caption{Feynman diagrams describing the $\tau \rightarrow \pi \nu_\tau$ decay.}
       \label{decay1}
\end{figure}

First, let us calculate the decay width  $\tau \to \pi \nu_{\tau}$ in the standard NJL model.
Here the Lagrangian  of the quark-meson interaction is
\ba
\mathcal {L} =\bar{q}[\tau^- \gamma_5(i g_{\pi}\pi^- + \gamma_{\mu} a_{1\mu}^-)]q
\label{LS}
\ea
In the standard NJL model the main contribution to the decay width $\tau \to \pi \nu_{\tau}$ is given by one of the quark loop
diagrams shown
in Fig.~\ref{decay1}.

Using the Lagrangian \eqref{LS} for the amplitude shown in Fig.~\ref{decay1} we obtain the corresponding expression:
\ba
A_{\tau \to \pi\nu_{\tau}}=\frac{G_F}{\sqrt{2}} \cdot \bar{u}_{\nu}\gamma_{\mu}\gamma_5 u_{\tau} \cdot g_{\mu \alpha} \cdot |V_{ud}|
\int\frac{d^4k}{(2\pi)^4}\frac{tr\,\, (i\gamma_5) (\hat {k} +m_u)\gamma_{\alpha} \gamma_5 (\hat {k}+\hat {p}+m_u)}{(m_u^2-k^2)(m_u^2-(k+p)^2)}.
\label{A}
\ea
Here $p$ is the momentum of the pion, $G_F = 1.16637 \cdot 10^{-11} MeV^{-2}$ is the Fermi constant,
$k$ is the momentum of the quark in the loop $k$, and $m_u$ is the $u$ quark mass, and
$|V_{ud}|$=0.97428 is the cosine of the Cabibbo angle.

We show that without the $\pi - a_1$ transitions this diagram completely determines the decay width  $\tau \rightarrow \pi \nu_\tau$.
Really, the amplitude \eqref{A} corresponding to this diagram is
\ba
A_{\tau \to \pi\nu_{\tau}} = L_{\mu} \cdot \frac{G_F}{\sqrt {2}}|V_{ud}| \biggl[\sqrt{2} g \cdot 4m_u I_2 =\sqrt{2} \frac{m_u}{g}= \sqrt{2} F_{\pi}\biggr]\cdot p^{\mu},
\label{In}
\ea
where $L_{\mu} = \bar{u}_{\nu}\gamma_{\mu}\gamma_5 u_{\tau}$ is the lepton current. \\
Here we used the following ratio:
$$4 I_2 = \frac{1}{g^2},$$ \\
where
$$F_{\pi} = 93\,\,MeV, \,\,m_u = m_d =280\,\,MeV.$$ \\
As a result, we get the well-known formula which can be found, for example, in \cite{Okun}. \\
When taking into account the $\pi - a_1$ transitions the coupling constant of the pion with the quarks $g_{\pi}$ has the form:
\ba
g_{\pi} = \sqrt{Z} g = \frac{m_u}{F_{\pi}},  \rm{where} \,\,\,\,Z=\biggl(1-\frac{6m_u^2}{m_{a_{1}}^2}\biggr)^{-1},
\ea
$m_{a_1}=1230\,\,\,MeV$  is the mass of the $a_1$ meson.

\subsection{The calculation of the decay with $\tau \to \pi \nu_{\tau}$ in the standard NJL model}

The decay amplitude of $\tau \to \pi (\pi') \nu_{\tau}$ has the form
\ba
F_{\tau \to \pi\nu_{\tau}} = L_{\mu} \cdot \frac{G_F}{\sqrt {2}}|V_{ud}||F_{1(\pi,\pi')}^q +F_{2(\pi,\pi')}^{a_1} +
F_{3(\pi,\pi')}^{a_1'}| p^{\mu}\cdot \pi (\pi'),
\label{FFF}
\ea
where $F_1^q$ corresponds to the contribution of the amplitude from the one-loop quark diagram (Fig.~\ref{decay1}).
$F_2^{a_1}$ corresponds to the contribution of the amplitude from the diagram with intermediate of the $a_1(1260)$ mesons (Fig.~\ref{decay2}).
$F_3^{a_1'}$ corresponds to the contribution of the amplitude from the diagram with intermediate of the $a_1'(1640)$ mesons (Fig.~\ref{decay2}).

The model parameters $F_{\pi}$ and $F_{\pi'}$ are determined by the equations

\ba
F=|F_1^q +F_2^{a_1} +F_3^{a_1'}|.
\label{FFF3}
\ea
In the standard NJL model for the quantities $F_i$ we obtain the following values
\ba
F_1^q =  Z F_{\pi}; \,\,\,\,\,\,F_2^{a_1} =(1-Z)F_{\pi},\,\,\,\,\,\,\,F_3^{a_1'} =0.
\ea
As a result, we get $F = F_{\pi}$.

It means that taking into account the $\pi - a_1$ transitions, we once again have the standard expression for the decay amplitude $\tau \to \pi\nu_{\tau}$.

The decay width of the $\tau \to \pi \nu_{\tau}$ process is equal to
\ba
\Gamma (\tau \to \pi \nu_{\tau}) = \frac{|F|^2}{2 \cdot 2 m_{\tau}}\Phi,
\label{Gamma}
\ea
where $\Phi$ is the phase volume:
\ba
\Phi = \frac{E_{\nu}}{4\pi m_{\tau}},
\ea
where $E_{\nu}$ is determined in the following form
\ba
E_{\nu} = \frac{m_{\tau}^2 - m_{\pi}^2}{2 m_{\tau}}.
\ea
We also used $(p_{\nu}p_{\tau}) = m_{\tau} E_{\nu}$.

For the square of the amplitude, we obtain the following expression
\ba
|F|^2 = 8 G_F^2|V_{ud}|^2 m_{\tau}^3 E_{\nu} F_{\pi}^2.
\label{M2}
\ea

This expression is in satisfactory agreement with the experimental data, $\Gamma^{theor}_{\tau \to \pi\nu_{\tau}} =2.5 \cdot 10^{-10}$ MeV, \,\,\,\,$\Gamma^{exp}_{\tau \to \pi\nu_{\tau}} =2.1 \cdot 10^{-10}$ MeV \cite{pdg}.

To describe the process $\tau \to \pi' \nu_{\tau}$ extended NJL model should be used, that will be done below.
To describe the process $\tau \to \pi \nu_{\tau}$, it is enough to use the standard NJL model.

\subsection{The decay $\tau \to \pi \nu_{\tau}$ in the extended NJL model}

Using the extended model we obtain only minor corrections to formulas from standard the NJL model.
Indeed, for example, we give the expression obtained in the extended model for the amplitude of the one-loop approximation
Fig.~\ref{decay1}).
\ba
F_1^q =  Z \biggl[F_{\pi}\frac{\sin(\alpha +\alpha_0)}{\sin(2\alpha_0)} + \Gamma \frac{\sin(\alpha -\alpha_0)}{\sin(2\alpha_0)}\biggr] p^{\mu}, \nn \\
F_2^{a_1} = D_1 +D_2 +D_3 +D_4, \,\,\,\,\, F_3^{a_1'} =C_1 +C_2 +C_3 +C_4,
\ea
where
\ba
\Gamma = \frac{I_2^{f}}{\sqrt{I_2 I_2^{ff}}} = 0.54.
\ea
Expressions $D_1 ,\,\,D_2,\,\, D_3,\,\, D_4$, and $C_1,\,\,C_2,\,\,C_3,\,\,C_4$, are shown in \eqref{D1} and \eqref{C1}, recpectively,
Hence it is easy to see that the contribution to the amplitude of the term field which contains the form factor that is negligibly small. \\
Therefore, for the calculation of the decay width $\tau \to \pi\nu_{\tau}$ one can use only the standard NJL model.
Since the angles $\alpha$ and $\alpha_0$ are close to each other, it is easy to see that part of the amplitude-form-factor provides a very small contribution, as compared to the first term, and it can be neglected.

\begin{figure}[!htb]
       \centering
	   \includegraphics[width=0.5\linewidth]{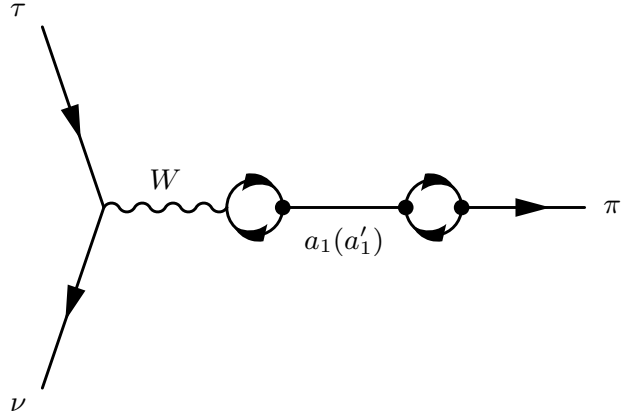}
       \caption{Diagrams describing the $\tau \to \pi \nu_{\tau}$ decay  with the intermediate axial vector mesons}
       \label{decay2}
\end{figure}

Now we consider the $\tau \to \pi\nu_{\tau}$  decay  which contains two quark-loop diagrams.
The corresponding quark loop diagram for this decay is shown in Fig.~\ref{decay2}).
First, we consider the diagram with the intermediate $a_1$ meson.
As a result we obtain
\ba
D_1 =-\frac{6m_u^2}{m_{a_1}^2} \sqrt{2} \biggl[\frac{\sin(\beta+\beta_0)}{\sin(2\beta_0)} + \Gamma \frac{\sin(\beta-\beta_0)}{\sin(2\beta_0)}\biggr] \cdot D_{F0} ,\nn \\
D_2 =-\frac{6m_u^2}{m_{a_1}^2}\frac{1}{g_{\rho}/2} \sqrt{2}\biggl[\frac{\sin(\beta+\beta_0)}{\sin(2\beta_0)} + \Gamma \frac{\sin(\beta-\beta_0)}{\sin(2\beta_0)}\biggr] \cdot D_{F1}, \nn \\
D_3=-\frac{6m_u^2}{m_{a_1}^2} \sqrt{2}\biggl[\frac{\sin(\beta+\beta_0)}{\sin(2\beta_0)} + \Gamma \frac{\sin(\beta-\beta_0)}{\sin(2\beta_0)}\biggr]\cdot D_{F11}, \nn \\
D_4=-\frac{6m_u^2}{m_{a_1}^2}\frac{1}{g_{\rho}/2} \sqrt{2}\biggl[\frac{\sin(\beta+\beta_0)}{\sin(2\beta_0)} + \Gamma \frac{\sin(\beta-\beta_0)}{\sin(2\beta_0)}\biggr] \cdot D_{F2},
\label{D1}
\ea
where $D_{F0}, D_{F1}, D_{F11}, D_{F2}$ correspond to the contribution from the second loop in Fig. ~\ref{decay2}.
They have the following form:
\ba
D_{F0} = Z F_{\pi}\biggl[\frac{\sin(\alpha+\alpha_0)}{\sin(2\alpha_0)} \cdot \frac{\sin(\beta+\beta_0)}{\sin(2\beta_0)}\biggr] \cdot p^{\alpha}, \nn \\
D_{F1} = \sqrt{\frac{3}{2}} \sqrt{Z} \Gamma F_{\pi}\biggl[\frac{\sin(\beta+\beta_0)}{\sin(2\beta_0)} \cdot \frac{\sin(\alpha-\alpha_0)}{\sin(2\alpha_0)}\biggr] \cdot p^{\alpha}, \nn \\
D_{F11} = \sqrt{Z} \Gamma F_{\pi}\biggl[\frac{\sin(\beta-\beta_0)}{\sin(2\beta_0)} \cdot \frac{\sin(\alpha+\alpha_0)}{\sin(2\alpha_0)}\biggr] \cdot p^{\alpha}, \nn \\
D_{F2} = \sqrt{\frac{3}{2}} F_{\pi}\biggl[\frac{\sin(\beta-\beta_0)}{\sin(2\beta_0)} \cdot \frac{\sin(\alpha-\alpha_0)}{\sin(2\alpha_0)}\biggr] \cdot p^{\alpha},
\label{DF}
\ea
We proceed to consider the contribution the diagrams with the intermediate $a_1'(1640)$ meson.
The amplitude comprising the $a_1'(1640)$ meson has the form:
\ba
C_1 =-\frac{6m_u^2}{m_{a_1'}^2}\biggl[\left(\frac{-\cos(\beta+\beta_0)}{\sin(2\beta_0)}\right) + \Gamma \left(\frac{-\cos(\beta-\beta_0)}{\sin(2\beta_0)}\right)\biggr] \cdot C_{F0}, \nn \\
C_2 =-\frac{6m_u^2}{m_{a_1'}^2}\frac{1}{g_{\rho}/2}\biggl[\left(\frac{-\cos(\beta+\beta_0)}{\sin(2\beta_0)}\right) + \Gamma \left(\frac{-\cos(\beta-\beta_0)}{\sin(2\beta_0)}\right)\biggr] \cdot C_{F1}, \nn \\
C_3=-\frac{6m_u^2}{m_{a_1'}^2}\biggl[\left(\frac{-\cos(\beta+\beta_0)}{\sin(2\beta_0)}\right) + \Gamma \left(\frac{-\cos(\beta-\beta_0)}{\sin(2\beta_0)}\right)\biggr] \cdot C_{F11}, \nn \\
C_4=-\frac{6m_u^2}{m_{a_1'}^2}\frac{1}{g_{\rho}/2}\biggl[\left(\frac{-\cos(\beta+\beta_0)}{\sin(2\beta_0)}\right) + \Gamma \left(\frac{-\cos(\beta-\beta_0)}{\sin(2\beta_0)}\right)\biggr] \cdot C_{F2}, \,\,\,\,
\label{C1}
\ea
where $C_{F0}, C_{F1}, C_{F11}, C_{F2}$ correspond to the contribution from the second loop in Fig. ~\ref{decay2}.
They have the following form:
\ba
C_{F0} = Z F_{\pi}\biggl[\frac{\cos(\beta+\beta_0)}{\sin(2\beta_0)} \cdot \frac{\sin(\alpha+\alpha_0)}{\sin(2\alpha_0)}\biggr] \cdot p^{\alpha}, \nn \\
C_{F1} = \sqrt{\frac{3}{2}} \sqrt{Z} \Gamma F_{\pi}\biggl[\frac{\cos(\beta+\beta_0)}{\sin(2\beta_0)} \cdot \frac{\sin(\alpha-\alpha_0)}{\sin(2\alpha_0)}\biggr] \cdot p^{\alpha}, \nn \\
C_{F11} = \sqrt{Z} \Gamma F_{\pi}\biggl[\frac{\cos(\beta-\beta_0)}{\sin(2\beta_0)} \cdot \frac{\sin(\alpha+\alpha_0)}{\sin(2\alpha_0)}\biggr] \cdot p^{\alpha}, \nn \\
C_{F2} = \sqrt{\frac{3}{2}} F_{\pi}\biggl[\frac{\cos(\beta-\beta_0)}{\sin(2\beta_0)} \cdot \frac{\sin(\alpha-\alpha_0)}{\sin(2\alpha_0)}\biggr] \cdot p^{\alpha}.
\label{CF}
\ea
As a result, after the calculation we get the following values:  \\
$F_1^q = 133.5 \,\,\,MeV,\,\,\,  F_2^{a_1} =-46.76 \,\,\,MeV, \,\,\,  F_3^{a_1'} = 5.9 \,\,\,MeV.$ \\
As a result, for $F_{\pi}$  we obtain $F_{\pi} = 92.8 \,\,\,MeV.$
It is easy to see that these results are in good agreement with each other.

\section{The decay $\tau \to \pi' \nu_{\tau}$ in the extended NJL model}

\begin{figure}[!htb]
       \centering
	   \includegraphics[width=0.5\linewidth]{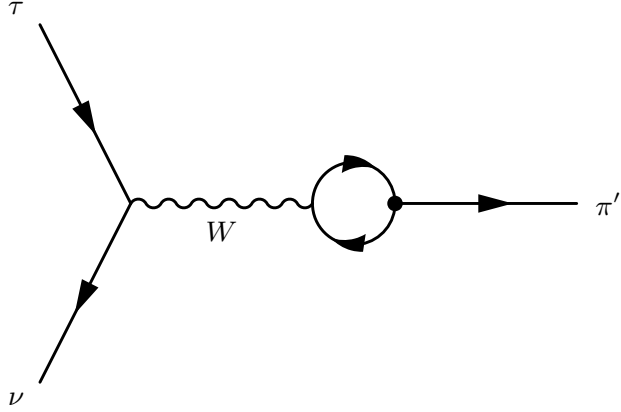}
       \caption{Feynman diagrams for the $\tau \rightarrow \pi' \nu_\tau$ decay.}
       \label{decay3}
\end{figure}

Now let us consider the decay of $\tau \to \pi' \nu_{\tau}$.
The corresponding quark diagram for this decay is shown in fig.~\ref{decay3}.

\begin{figure}[!htb]
       \centering
	   \includegraphics[width=0.5\linewidth]{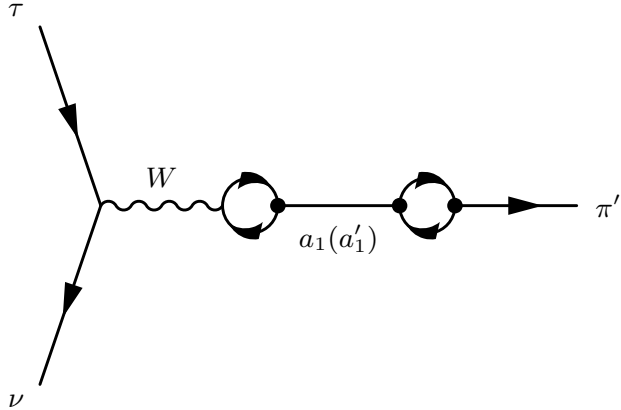}
       \caption{Feynman diagrams for the $\tau \to \pi' \nu_{\tau}$ decay  with the axial vector transitions.}
       \label{decay4}
\end{figure}

Note that in the vertex with $\pi'$ there are two cases: in the first case, this vertex does not contain the vertex form factor,
and in the second case, this vertex does.
As the result, this amplitude takes the following form:
\ba
A_0 = L_{\mu} \cdot \frac{G_F}{\sqrt {2}}|V_{ud}| \cdot \sqrt {2} \sqrt {Z} F_{\pi} \biggl[\sqrt{Z} \frac{\cos(\alpha +\alpha_0)}{\sin(2\alpha_0)} +
\Gamma \cdot \frac{\cos(\alpha-\alpha_0)}{\sin(2\alpha_0)}\biggr]p^{\mu} = \nn \\
L_{\mu} \cdot \frac{G_F}{\sqrt {2}}|V_{ud}| \cdot (-5.64) p^{\mu}\,\,\,MeV.
\label{A0}
\ea
Besides the one-loop contribution to the decay $\tau \to \pi' \nu_{\tau}$ an important role is played by the
contributions of the intermediate $a_1 (1260)$ and $a_1' (1640)$ axial vector mesons.
The corresponding two-loop diagrams for this contribution are shown in Fig.~\ref{decay4}.
Here the first loop consists of two parts one of which does not contain a form factor at the vertices with the $a_1$ ($a_1'$) mesons.
The second loop of each of the vertices consists of two parts, one of which comprises a form factor and the other does not.

We first consider the diagrams with the intermediate $a_1$ meson. \\
As the result, we obtain:
\ba
A_1 =-6m_u^2 \sqrt{2} \biggl[\frac{\sin(\beta+\beta_0)}{\sin(2\beta_0)} + \Gamma \frac{\sin(\beta-\beta_0)}{\sin(2\beta_0)}\biggr]\cdot BW \cdot A_{F0} ,\nn \\
A_2 =-6m_u^2\frac{1}{g_{\rho}/2} \sqrt{2}\biggl[\frac{\sin(\beta+\beta_0)}{\sin(2\beta_0)} + \Gamma \frac{\sin(\beta-\beta_0)}{\sin(2\beta_0)}\biggr]\cdot BW \cdot A_{F1}, \nn \\
A_3=-6m_u^2 \sqrt{2}\biggl[\frac{\sin(\beta+\beta_0)}{\sin(2\beta_0)} + \Gamma \frac{\sin(\beta-\beta_0)}{\sin(2\beta_0)}\biggr]\cdot BW \cdot A_{F11}, \nn \\
A_4=-6m_u^2\frac{1}{g_{\rho}/2} \sqrt{2}\biggl[\frac{\sin(\beta+\beta_0)}{\sin(2\beta_0)} + \Gamma \frac{\sin(\beta-\beta_0)}{\sin(2\beta_0)}\biggr]\cdot BW \cdot A_{F2},
\label{A1}
\ea
where $A_{F0}, A_{F1}, A_{F11}, A_{F2}$ correspond to the contribution from the second loop in Fig.~\ref{decay4}.
They have the following form:
\ba
A_{F0} = Z F_{\pi}\biggl[\frac{\sin(\beta+\beta_0)}{\sin(2\beta_0)} \cdot \biggl(-\frac{\cos(\alpha+\alpha_0)}{\sin(2\alpha_0)}\biggr)\biggr] \cdot p^{\alpha}, \nn \\
A_{F1} = \sqrt{\frac{3}{2}} \sqrt{Z} \Gamma F_{\pi}\biggl[\frac{\sin(\beta-\beta_0)}{\sin(2\beta_0)} \cdot \biggl(-\frac{\cos(\alpha+\alpha_0)}{\sin(2\alpha_0)}\biggr)\biggr] \cdot p^{\alpha}, \nn \\
A_{F11} = \sqrt{Z} \Gamma F_{\pi}\biggl[\frac{\sin(\beta+\beta_0)}{\sin(2\beta_0)} \cdot \biggl(-\frac{\cos(\alpha-\alpha_0)}{\sin(2\alpha_0)}\biggr)\biggr] \cdot p^{\alpha}, \nn \\
A_{F2} = \sqrt{\frac{3}{2}} F_{\pi}\biggl[\frac{\sin(\beta-\beta_0)}{\sin(2\beta_0)} \cdot \biggl(-\frac{\cos(\alpha-\alpha_0)}{\sin(2\alpha_0)}\biggr)\biggr] \cdot p^{\alpha},
\label{AF}
\ea
We proceed to consider the contribution from the diagrams with the intermediate $a_1'(1640)$ mesons.
The amplitude containing the $a_1'(1640)$ meson has the form:
\ba
B_1 =-6m_u^2 \sqrt{2} \biggl[\left(\frac{-\cos(\beta+\beta_0)}{\sin(2\beta_0)}\right) + \Gamma \left(\frac{-\cos(\beta-\beta_0)}{\sin(2\beta_0)}\right)\biggr]\cdot BW \cdot B_{F0}, \nn \\
B_2 =-6m_u^2\frac{1}{g_{\rho}/2} \sqrt{2} \biggl[\left(\frac{-\cos(\beta+\beta_0)}{\sin(2\beta_0)}\right) + \Gamma \left(\frac{-\cos(\beta-\beta_0)}{\sin(2\beta_0)}\right)\biggr]\cdot BW \cdot B_{F1}, \nn \\
B_3=-6m_u^2 \sqrt{2} \biggl[\left(\frac{-\cos(\beta+\beta_0)}{\sin(2\beta_0)}\right) + \Gamma \left(\frac{-\cos(\beta-\beta_0)}{\sin(2\beta_0)}\right)\biggr]\cdot BW \cdot B_{F11}, \nn \\
B_4=-6m_u^2\frac{1}{g_{\rho}/2} \sqrt{2} \biggl[\left(\frac{-\cos(\beta+\beta_0)}{\sin(2\beta_0)}\right) + \Gamma \left(\frac{-\cos(\beta-\beta_0)}{\sin(2\beta_0)}\right)\biggr]\cdot BW \cdot B_{F2}, \,\,\,\,
\label{B1}
\ea
where $B_{F0}, B_{F1}, B_{F11}, B_{F2}$ correspond to the contributions from the second loop in Fig.~\ref{decay4}.
They have the following form:
\ba
B_{F0} = Z F_{\pi}\biggl[\left(-\frac{\sin(\beta+\beta_0)}{\sin(2\beta_0)}\right) \cdot \biggl(-\frac{\cos(\alpha+\alpha_0)}{\sin(2\alpha_0)}\biggr)\biggr] \cdot p^{\alpha}, \nn \\
B_{F1} = \sqrt{\frac{3}{2}} \sqrt{Z} \Gamma F_{\pi}\biggl[\left(-\frac{\cos(\beta-\beta_0)}{\sin(2\beta_0)}\right) \cdot \biggl(-\frac{\cos(\alpha+\alpha_0)}{\sin(2\alpha_0)}\biggr)\biggr] \cdot p^{\alpha}, \nn \\
B_{F11} = \sqrt{Z} \Gamma F_{\pi}\biggl[\left(-\frac{\cos(\beta+\beta_0)}{\sin(2\beta_0)}\right) \cdot \biggl(-\frac{\cos(\alpha-\alpha_0)}{\sin(2\alpha_0)}\biggr)\biggr] \cdot p^{\alpha}, \nn \\
B_{F2} = \sqrt{\frac{3}{2}} F_{\pi}\biggl[\left(-\frac{\cos(\beta-\beta_0)}{\sin(2\beta_0)}\right) \cdot \biggl(-\frac{\cos(\alpha-\alpha_0)}{\sin(2\alpha_0)}\biggr)\biggr] \cdot p^{\alpha},
\label{BF}
\ea
where $BW$ is the Breit-Wigner formula for the intermediate state of the $a_1 (a_1')$ mesons and has the form:
\ba
BW = \frac{1}{M^2_{a_1 (a_1')}-M^2_{\pi'} -\mathrm{i} M_{a_1(a_1')}\Gamma_{a_1(a_1')}}.
\ea
For the  $\tau \to \pi' \nu_{\tau}$ decay in the extended NJL model we obtain the following value for the the quantities
$F_{1(\pi')}^q,\,\,\,\,  F_{2(\pi')}^{a_1},\,\,\,\,   F_{3(\pi')}^{a_1'}$
\ba
F_{1(\pi')}^q = 5.64, \,\,\,\,F_{2(\pi')}^{a_1} = -2.3 - i \cdot 6.39, \,\,\,\,F_{3(\pi')}^{a_1'} =-4.59 + i \cdot 1.88.
\ea
\ba
\tilde{F_{\pi'}} = F_{1(\pi')}^q + F_{2(\pi')}^{a_1} + F_{3(\pi')}^{a_1'} =-1.25 - i \cdot 4.51, \nn \\
F_{\pi'} = |\tilde{F_{\pi'}}| = \sqrt {Re \tilde{F_{\pi'}}^2 + Im \tilde{F_{\pi'}}^2} = 4.68\,\,\,MeV.
\ea
Calculations on the lattice gave the following results for $F_{\pi'} \approx 2.4 \,\,\,MeV$ \cite{Kim}.
We can see that our results are in qualitative agreement with the estimates obtained by calculations on the lattice

As the result, the total decay amplitude $\tau \to \pi' \nu_{\tau}$ takes the form:
\ba
T=-A_0+A_1 + A_2 +A_3 +A_4 +B_1 +B_2 +B_3 +B_4.
\label{M}
\ea

Considering the decay amplitude $\tau \to \pi' \nu_{\tau}$ via one loop \eqref{A0} for the decay width \eqref{Gamma}
we obtain the value of $\Gamma_{\tau \to \pi' \nu_{\tau}} = 1.04 \cdot 10^{-13}$ MeV.
For the decay width via two loops with the intermediate $a_1 (1260)$ mesons we obtain the value of
$\Gamma_{\tau \to (a_1) \to\pi' \nu_{\tau}} = 3.03 \cdot 10^{-13}$ MeV.
With the intermediate $a_1' (1640)$ mesons we obtain the value of $\Gamma_{\tau \to (a_1') \to\pi' \nu_{\tau}} = 1.62 \cdot 10^{-13}$ MeV.
The total decay width of $\tau \to \pi' \nu_{\tau}$ with the interference between the amplitudes \eqref{M} is
$\Gamma_{\tau \to \pi' \nu_{\tau}}^{total} = 2.21 \cdot 10^{-13}$ MeV.

The resulting value for the decay width of $\tau \to \pi' \nu_{\tau}$ is in satisfactory agreement with the experimental evaluation
obtained in \cite{exp}.
There, the following restrictions on the quantity of the decay width $\tau \to \pi' \nu_{\tau}$ was obtained that is,
$\Gamma_{\tau \to (\pi')\nu_{\tau}}^{exp} = 2.27 \cdot 10^{-13} MeV \div 4.31 \cdot 10^{-13} MeV$.

In addition, the evaluation obtained for $F_{\pi'} = 4.68\,\,\,MeV$ is in qualitative agreement with the estimates obtained in the lattice model
\cite{Kim}.

\section{conclusion}

The decay mechanism $\tau \to \pi \nu_{\tau}$ is similar to the decay $\pi^- \to \mu^- \nu$, which was described in \cite{volkov9}.
Those calculations were performed in the standard NJL model.
It is interesting to note that in the calculation of these processes in the quark one-loop approximation with the decay constant $F_{\pi}$
there appears a factor $Z$ that is due to taking into account of $\pi - a_1$ transitions.
To obtain a correct form for the decay amplitude $\tau \to \pi \nu_{\tau}$, it is also necessary to take into account the two-loop
approximation with the intermediate $a_1(1260)$ mesons.
This approach takes into account the possibility of the $a_1 - \pi$ transition to the intermediate state with the $a_1$ meson.
As a result, for the same value of $F_{\pi}$=93 MeV we obtain satisfactory agreement with the experimental data in both these cases,
as the decay $\tau \to \pi \nu_{\tau}$, and the decay $\pi^- \to \mu^- \nu$ \eqref{FFF3}.

It is interesting to note that the calculation of the width $\tau \to \pi \nu_{\tau}$ conducted in the expanded NJL model taking into account the intermediate $a_1(1260)$ and $a_1'(1640)$  leads to the same results for the width of the $\tau \to \pi \nu_{\tau}$ decay and of the constant $F_{\pi}$.
The extended NJL model allows us to describe the decay $\tau \to \pi' \nu_{\tau}$.

Unlike  the case of the decay from $\tau \to \pi \nu_{\tau}$ in the decay $\tau \to \pi' \nu_{\tau}$ amplitude the intermediate
$a_1(1260)$ and $a_1'(1640)$ mesons play a decisive role.

Note that the results obtained here for the widths of the $\tau \to \pi' \nu_{\tau}$ decay is in satisfactory agreement with the experimental data
\cite{exp}.
In our calculation, we can claim only qualitative agreement with experiment, as in the extended NJL model to describe the excited state of the axial-vector mesons we used the mixing angles taken from the vector meson sector \cite{volkov10}.

\section{Acknowledgements}

The authors are grateful to A.B.Arbuzov and D.G.Kostunin for fruitful discussions and useful comments.

\end{document}